\documentclass[aps,prl,preprintnumbers,a4,showpacs,twocolumn,nofootinbib]{revtex4}

\newcommand{\be}{\begin{eqnarray}}
\newcommand{\ee}{\end{eqnarray}}

\def\hbar#1{\slash\hspace{-2.5mm}#1}
\usepackage{graphicx,graphics}
\usepackage{mathrsfs}
\usepackage{amssymb,amsmath}
\usepackage{dcolumn}
\usepackage{bm}
\usepackage{slashed}
\usepackage{color}

\newcommand{\beq}[1] {\begin{equation}\label{#1} }
\newcommand{\eeq} {\end{equation} }
\newcommand{\bea}[1]{\begin{eqnarray}\label{#1} }
\newcommand{\eea}{\end{eqnarray}}

\begin{document}

\preprint{OSU-HEP-11-06}
\preprint{CUMQ/HEP 70}
\title{Top quark asymmetry and $Wjj$ excess at CDF from gauged flavor symmetry}
\author{K.S.~Babu$^a$}
\author{Mariana Frank$^b$}
\author{Santosh Kumar Rai$^a$}
\affiliation{$^a$Department of Physics and Oklahoma Center for High Energy Physics,
Oklahoma State University, Stillwater, Oklahoma, USA 74078\\
$^b$Department of Physics, Concordia University,
7141 Sherbrooke St. West, Montreal, Quebec, Canada H4B 1R6}

\begin{abstract}
We show that the scalar sector needed for fermion mass generation when the flavor symmetry
of the standard model is maximally gauged can consistently explain two anomalies reported
recently by the CDF collaboration -- the forward-backward asymmetry in $t\overline{t}$ pair
production, and the dijet invariant mass in the $Wjj$ channel.  A pair of nearly degenerate scalar
doublets with masses in the range 150-200 GeV explain these anomalies, with additional
scalars predicted in the mass range 100-400 GeV.  Consistency of such low scale flavor physics
with flavor changing processes is shown, and expectations for the LHC are outlined.
\end{abstract}
\maketitle

{\it Introduction:}  The well--established gauge sector of the standard $SU(3)_C \times SU(2)_L \times U(1)_Y$ model (SM)
with three families of quarks and leptons possesses a $[U(3)]^5$ global flavor symmetry, with a separate
$U(3)$ assigned to every fermion type. Here we pursue gauging a maximal subgroup of
this symmetry, following the dictum that any symmetry that is anomaly-free must be gauged.  This will provide
an organizing principle for the fermions and will also shed light on the origin of their masses and mixings.
The maximal subgroups of $[U(3)]^5$ that can be gauged are found to be: (A) $O(3)_{L\,{\{Q, L\}}} \times O(3)_{R\,{\{u^c, d^c, e^c\}}}$,
(B) $O(3)_{\{Q,u^c,e^c\}} \times O(3)_{\{L,d^c\}}$
and (C) $SU(3)_{\{Q,u^c,d^c\}} \times O(3)_{\{L,e^c\}}$. This result follows from anomaly cancelation,
especially the $G^3$  and the mixed $G^2 \times Y$ anomalies, where $G$ is the gauged flavor symmetry.
Case (A) appears to us to be most promising, with the left--handed $SU(2)_L$ doublet fermions $(Q,L)$
being triplets of $O(3)_L$ and the (conjugate) right--handed singlet fermions $(u^c,d^c,e^c)$ being triplets of $O(3)_R$.  The
SM Higgs doublet fields $\Phi^a$ that generate fermion masses and mixings must then transform as $(3,3)$ of $O(3)_L \times
O(3)_R$.  The proliferous and disparate Yukawa couplings of the SM are thus promoted to dynamical fields $\Phi^a_{ij}$,
with a single unified coupling for each fermion type.
This setup, as we show below, provides a unified and natural explanation of two anomalies reported by the CDF collaboration recently,
in the $t\overline{t}$ forward backward asymmetry, and in the dijet mass distribution in the $Wjj$
final state.

The CDF collaboration has reported a parton-level forward-backward (FB) asymmetry in the top quark pair
production in the $t\bar{t}$ rest frame to be $A_{t\overline{t}}=0.475 \pm 0.114$ \cite{Aaltonen:2011kc}
for large $t\bar{t}$ invariant mass, $M_{t\bar{t}}>450$ GeV, which is
3.4 standard deviations above the NLO QCD prediction of
$0.088 \pm 0.013$.
The same collaboration has also reported an excess in the 120-160 GeV
mass range for dijet invariant mass in the $Wjj$ channel \cite{Aaltonen:2011mk},
which is 3.2 standard deviations away from SM predictions. This result is consistent with the
production and decay of a new particle with a mass of $\sim 150$ GeV into two jets in association with a $W$.
No excess is seen in the $W b b$, $W\ell \ell$, or the $Zjj$ channels.  These two anomalies appear to be significant
enough to hint at new physics beyond the SM. We suggest that the $\Phi^u(3,3)$ scalar fields of the $O(3)_L \times O(3)_R$ flavor symmetric model,
responsible for the up--type quark mass generation, provides the needed new physics, consistent with the associated constraints.

{\it Model:}  We extend the gauge symmetry of the SM to contain $O(3)_L \times O(3)_R$, a maximal subgroup of the flavor symmetry,
with the three families of fermions
assigned as $\{Q(3,1) + L(3,1) + u^c(1,3) + d^c(1,3)+e^c(1,3)\}$ under the flavor group. This renormalizable theory, while maximally symmetric,
is rather minimal in the sense that no new fermions are introduced.
The Higgs sector for fermion mass generation consists of $\Phi^u(3,3)$
and $\Phi^d(3,3)$ which are SM doublets with $Y=\pm\frac{1}{2}$.  Two such fields are needed in order to avoid the proportionality
relations $m_u:m_c:m_t = m_d:m_s:m_b$.  Fermion masses arise from  the Yukawa Lagrangian
\begin{equation}
\label{Yuk}
{\cal L}_{\rm Yuk} = Y_u Q_i u^c_j \Phi^u_{ij} + Y_d Q_i d^c_j \Phi^d_{ij} + Y_\ell L_i e^c_j \Phi^d_{ij} + h.c.
\end{equation}
where $i,j = 1-3$ are $O(3)_{L,R}$ indices. For simplicity
we have assumed a discrete $Z_2$ symmetry under which $d^c_j, e^c_j$ and $\Phi^d_{ij}$ are odd. The quark mass matrices
following from Eq. (\ref{Yuk}) have elements $M^{u,d}_{ij} = Y_{u,d} \langle \Phi^{u,d}_{ij}\rangle$, with a single unified Yukawa
coupling in each sector.  The observed mass hierarchy among the quarks
is explained dynamically with a hierarchical structure in the vacuum expectation values (VEV) $\langle \Phi^{u,d}_{ij}\rangle$
(and similarly for the leptons \cite{fn}). Realistic quark mixings will also be induced by (\ref{Yuk}).
It follows from (\ref{Yuk}) that $Y_u \simeq m_t/v_u$, $Y_d \simeq m_b/v_d$, $Y_\ell \simeq m_\tau/v_d$, where $v_{u,d} \equiv \langle \Phi_{33}^{u,d}\rangle$ with $v_u^2+v_d^2 \simeq (174 ~{\rm GeV})^2$. In particular, for $v_u \simeq v_d$, we have
$Y_u \simeq 1.4$, with $Y_d,Y_\ell \ll Y_u$.

If the $\Phi^{u}$ scalars are to explain the CDF anomalies, some of its components should have masses in the 150-200 GeV range.
An immediate question is whether such light scalars, with ${\cal O}(1)$  Yukawa couplings to the $u,c,t$ quarks,
are compatible with flavor changing constraints.  The naive expectation that scalar and vector bosons associated with flavor physics
must have masses in excess of ${\cal O} (100)$ TeV does not hold in the present model, owing to approximate symmetries.
If the scalar fields
$\Phi^a_{ij}$ are nearly mass eigenstates, and if the right--handed fermion mixing angles (which are unphysical in the SM) are small,
these scalars can be relatively light.  For example, the neutral $\Phi^u_{12}$ will induce an effective operator $|Y_u|^2(\overline{u}_Lc_R)
(\overline{c}_Ru_L)/M_{\Phi^u_{12}}^2$, which does not lead to $D^0-\overline{D^0}$ mixing, even allowing for left--handed quark mixings.
While the $O(3)_L$ gauge bosons must be heavier than about 30 TeV (from a combination of $K^0-\overline{K^0}$ and $D^0-\overline{D^0}$ mixing
constraints), the $O(3)_R$ gauge bosons can be relatively light, although in the present paper we will take them to be beyond
the reach of Tevatron  \cite{pierce}.

We assume that the $O(3)_L \times O(3)_R$ flavor gauge symmetry is spontaneously broken at an energy scale much above the weak scale
by SM singlet scalar fields $T_L(7,1)+T_R(1,7)$.  These scalars leave behind an unbroken discrete subgroup,
ensuring that $\Phi^a_{ij}$ will be near mass eigenstates.  When the 7--plet of $O(3)$, which is a symmetric traceless tensor $T^{ijk}$
(with $T^{ijj}=0$, $i,j,k =1-3$) acquires a VEV along $T_{L,R}^{111} = -T_{L,R}^{122} = V_{L,R}$, the  $O(3)_L \times O(3)_R$ symmetry breaks to
the discrete group $Q_6 \times Q_6$ \cite{kephart}.  Under this unbroken symmetry \cite{bk} $\Phi^{u,d}$ will transform as $(1,1) + (1,2) + (2,1) + (2,2)$,
which guarantees the absence of mixing between various states.  From the Higgs potential couplings of $\Phi^a$ with $T_{L,R}$
given by
{\small
\begin{eqnarray}
V \supset \kappa_{1L}^a T_L^{ijk}T_L^{ijk} \,{\rm Tr}(\Phi^{a \dagger} \Phi^a) + \frac{\kappa_{2L}^a}{4} (\Phi^a \Phi^{a\dagger})^{ij} T_L^{ikl}T_L^{jkl}
\nonumber \\
+ \kappa_{1R}^a T_R^{ijk}T_R^{ijk}\, {\rm Tr}(\Phi^{a \dagger} \Phi^a) + \frac{\kappa_{2R}^a}{4} (\Phi^{a\dagger} \Phi^a)^{ij} T_R^{ikl}T_R^{jkl}
\end{eqnarray}}
we obtain the mass relations
{\small
\begin{eqnarray}
\label{masses}
m^2_{\{\Phi_{12}^u,\Phi_{13}^u\}} = \mu_u^2+\kappa_{2L}^uV_L^2; ~~m^2_{\{\Phi_{21}^u,\Phi_{31}^u\}} = \mu_u^2+\kappa_{2R}^uV_R^2; \nonumber \\
m^2_{\Phi^u_{11}} = \mu_u^2 + \kappa_{2L}^u V_L^2 + \kappa_{2R}^u V_R^2; ~~
m^2_{\{\Phi_{22}^u,\Phi_{23}^u,\Phi_{32}^u,\Phi_{33}^u \}} = \mu_u^2
\end{eqnarray}}
\hspace*{-0.3cm} where $\mu_u^2$ is an effective mass parameter, with similar results for $m^2_{\Phi^d_{ij}}$.  We shall identify the degenerate pair
$\Phi^u_{12}$ and $\Phi_{13}^u$ with masses in the 150-200 GeV as being responsible for the CDF dijet excess and the
$t\overline{t}$ asymmetry respectively.  Since $\langle\Phi^u_{33}\rangle = v_u \sim 100$ GeV, this field must have a negative
squared mass. It follows from Eq. (\ref{masses}) that the entire $(2,2)$ component of $Q_6 \times Q_6$ from $\Phi^u$ must have masses
comparable to $\Phi_{12}^u$.  Quartic self--couplings of
$\Phi^u$ will induce mass splitting among these fields, as well as between the neutral and the charged members of the same
doublets, and will provide positive squared masses for all the fields.  These corrections cannot make the
$\Phi^u(2,2)$ much above 400 GeV or so, which is a prediction of the model correlated with the CDF anomaly.  Note that the other
components of $\Phi^{u,d}$ have independent masses and can be pushed beyond the Tevatron reach.

The Lagrangian that couples the scalars $\Phi_{12}^u, \Phi_{13}^u$ to the fermions is given explicitly as
{\small
\begin{equation}
{\cal L} = Y_u [\overline{u}_L c_R \Phi_{12}^0 + \overline{d}_L c_R \Phi_{12}^- + \overline{u}_L t_R \Phi_{13}^0
+\overline{d}_L t_R \Phi_{13}^-] + h.c.
\label{yuk1}
\end{equation}
}These scalars also couple to the SM gauge bosons ($\gamma, Z,W^\pm$) via
the kinetic terms $(D^\mu\Phi_{1j})^\dagger (D_\mu\Phi_{1j})$, where $j=2,3$.
Eq. (4) will lead to new contributions to top quark pair
production as well as the dijet searches at the Tevatron
and the LHC \cite{nelson}.  The new scalars will be produced on-shell if
kinematically accessible and will decay hadronically.
To account for the dijet bump seen in the CDF experiment in the
$Wjj$ channel we choose the mass of the neutral scalars, $m_{\Phi_{1j}^0}=150$ GeV ($j=2,3$)
while the charged scalars have a mass of $m_{\Phi_{1j}^\pm}=180$ GeV.
(We assume that the scalar and pseudoscalars from $\Phi_{1j}^0$ are degenerate.)
This choice also explains the FB asymmetry in the top
pair production. $\Phi_{12}$ causes the dijet anomaly, while $\Phi_{13}$
causes the top quark asymmetry.  No $Wbb$ or $W \ell \ell$ events are expected, since
$\Phi_{12}$ has no couplings (or has highly suppressed couplings) to $b$ quarks or leptons.  Equality of the $\Phi_{12}$ and $\Phi_{13}$ masses
follows from the $Q_6 \times Q_6$ symmetry.  The Yukawa coupling strength of $Y_u=1.4$ is found to be the
favorable choice with both observed phenomena at CDF.
The new scalars contribute to the electroweak T-parameter, which is 0.032 for
our choice of masses, consistent with precision electroweak data.

Such light scalars $\Phi_{1j}$ which couple strongly to
quarks would lead to large resonant dijet cross sections at hadron machines.
While these effects are swamped by QCD background at the Tevatron, the UA2 collaboration
at the CERN SPS collider has placed bounds on dijet resonances \cite{Alitti:1993pn}.
For scalar masses of 150 GeV (180 GeV) 
the 90\% C.L. upper bounds are $\approx$ 90 pb (50 pb).
The resonant cross sections for our case are $\approx$ 70 pb (16 pb),
well within these limits. We also note that for our choice of scalar masses
the decay $t \rightarrow u \Phi_{13}^{0*}$ is open and modifies
the top quark width to $\Gamma_t \simeq 1.72$ GeV, which is consistent with direct
measurements at Tevatron \cite{Abazov:2010tm}.

\begin{center}
\begin{figure}[!ht]
\includegraphics[width=3.1in,height=2.3in]{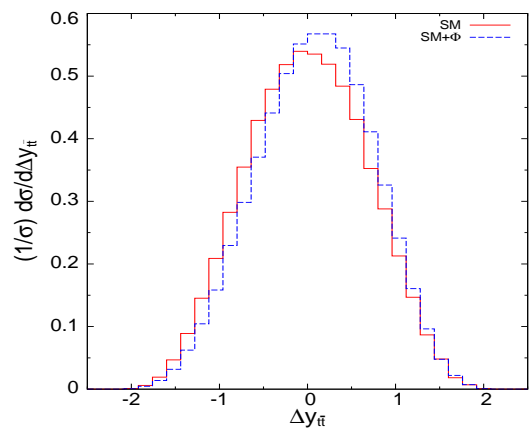}
\caption{The normalized differential cross section as a function of the rapidity difference $\Delta y$ for our model and SM at LO in the $t\bar{t}$ rest frame.}
\label{Att1}
\end{figure}
\end{center}

\vspace*{-0.3in}
{\it Top quark forward-backward asymmetry:}
The recent CDF result for a large $t\bar{t}$ FB asymmetry is more than $3\sigma$
away from the SM NLO predictions \cite{Aaltonen:2011kc}.
The asymmetry also shows a distinct mass dependence and is more pronounced
for larger values of the top-pair invariant mass. No new effect
is however transparent in the total $t\bar{t}$ cross section or its differential distribution
in the invariant mass. Any new physics should then interfere destructively
with the SM s-channel production modes. In our model, this occurs through the
t-channel contribution via $\Phi_{13}^0$ and $\Phi_{13}^\pm$ exchange in the subprocesses
$u\bar{u}\to t\bar{t}$ and $d\bar{d}\to t\bar{t}$.
We find that the for $Y_u \simeq 1.4$, significant FB asymmetry is generated, while
the $t\bar{t}$ cross section  changes by less that 5\% from the SM.
The cross section does show a slight dependence on the large invariant mass of the top
pairs ($> 450$ GeV) but is within 10\% of the SM values, which should be consistent with data.

\begin{center}
\begin{figure}[!ht]
\includegraphics[width=3.1in,height=2.3in]{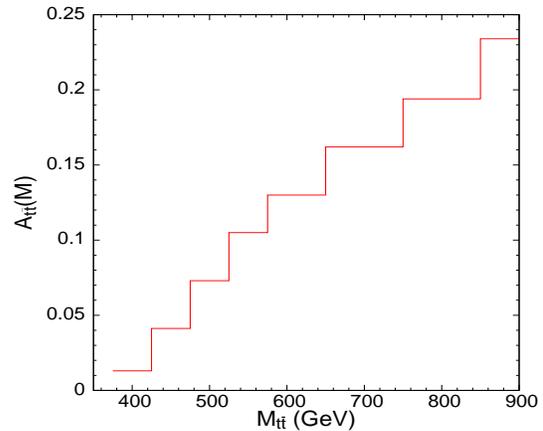}
\caption{The asymmetry $A_{t\bar{t}}$ is shown as a function of $M_{t\bar{t}}$, in the $t\bar{t}$ rest frame.}
\label{Att2}
\end{figure}
\end{center}

To show our results we do a parton level calculation by implementing our model in
{\tt CalcHEP} \cite{Pukhov:2004ca}. We use the rapidity coverage for the top quark to be
$|\eta|<2.0$ and construct the asymmetry in the $t\bar{t}$ rest frame defined by
\begin{align}
A_{t\overline{t}} = \frac{N_{\Delta y >0}-N_{\Delta y < 0}}{N_{\Delta y > 0}+N_{\Delta y < 0}}
\end{align}
where $\Delta y$ is the frame independent rapidity difference $(y_t-y_{\bar{t}})$. In Fig. \ref{Att1}
we show the differential cross section against the rapidity difference $\Delta y$ for our model as well
as the LO SM contribution (which is symmetric about $\Delta y=0$). An asymmetry
is evident in the distribution with the new scalars contributing to the $t\bar{t}$
production. We find an asymmetry of $0.104$ with no selection condition on the invariant
mass of $t\bar{t}$ which when combined with SM NLO result ($0.058\pm 0.009$) is in close
agreement with the observed value of $0.158\pm 0.075$. CDF reports a larger deviation ($>3\sigma$) for
events with $M_{t\bar{t}}>450$ GeV. In our model we get $A_{t\overline{t}}\simeq 0.156$ which when combined
with the SM value of $0.088\pm 0.013$  is within two sigma of the observed CDF value
of $0.475 \pm 0.114$. We also compute the asymmetry $A_{t\bar{t}}(|\Delta y|\geq 1)$ which in our
model is $\sim 0.195$ and when combined with the SM value of $0.123\pm 0.008$ is within one sigma
of the observed CDF value of $0.611\pm0.256$. In Fig. \ref{Att2} we show the asymmetry as a function
of $M_{t\bar{t}}$ in bins of 50 GeV in the range of 350-600 GeV and subsequent
bins of 100 GeV, which is in agreement with the behavior observed by CDF.

\vspace*{-0.3in}
{\it Excess in the $Wjj$ events:}
Our model predicts a doublet of scalars $\Phi_{12}$ nearly degenerate with $\Phi_{13}$
that couples to the $(u,d)$ and the charm quark
as given in (\ref{yuk1}). Although the charm quark flux is suppressed compared to the
valence $u,d$ quarks in the proton PDFs, with  $Y_u \simeq 1.4$ it
leads to the ccorrect magnitude for the cross section needed to explain the $Wjj$ excess at CDF \cite{Aaltonen:2011mk}.

\begin{center}
\begin{figure}[!ht]
\includegraphics[width=3.1in,height=2.3in]{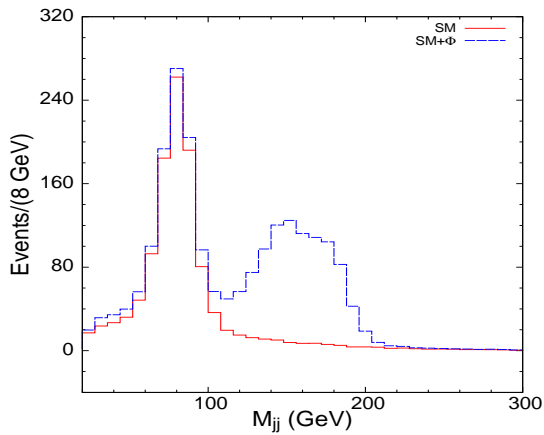}
\caption{The dijet invariant mass distribution in our model and in the SM $WW$ and $ZZ$ channels.}
\label{Wjj}
\end{figure}
\end{center}

The cross sections for the subprocesses that contribute to the production of a  single
scalar with associated $W$ boson at LO
in our model are: $\sigma(c \overline{d} \rightarrow W^+ \Phi_{12}^{0*}) = 765~{\rm fb}$, 
$\sigma(u \overline{c} \rightarrow W^+ \Phi_{12}^-) = 919~{\rm fb}$, and their charge conjugates.
This adds up to a total excess cross section for the $Wjj$ final state at LO of $\sim 3.37$ pb.
Taking into account K-factors of 1.3 \cite{Brein:2004ue} would give us a cross section of about 4.38 pb.
For our analysis of the dijet events with only the $WW$ and $WZ$ contribution retained for the SM,
we use the {\tt CalcHEP} generated events for production and decay and pass it through a {\tt Pythia} \cite{Sjostrand:2006za} interface which implements the showering and hadronization of the events.
For the jet construction, we use the inbuilt jet-clustering algorithm PYCELL
in {\tt Pythia} with CDF detector parameters. We implement the cuts on our final
state configuration of $1\ell + 2j + \slashed{E}_T + X$ where $\ell=e,\mu$ as given
by the CDF analysis. The charged lepton must have $p_T>20$ GeV and be within the rapidity
gap $|\eta|<1.0$. The events must have a minimum missing transverse momentum
$\slashed{E}_T>25$ GeV, only 2 jets, each with a $p_T>30$ GeV and $|\eta|<2.4$ and
the dijet system must have $p_T>40$ GeV. The jets with a charged lepton in a cone
of $\Delta R=0.52$ are rejected. We also implement the transverse mass condition on
the lepton+$\slashed{E}_T > 30$ GeV. The $WW$ and $WZ$ SM background were generated
using {\tt Pythia} and passed through the same set of kinematic cuts.
In Fig. \ref{Wjj} we show the invariant mass distribution of the dijet system which
shows the distinct bump beyond the SM weak boson resonances. The cross sections
for the signal and SM background have been multiplied by respective K-factors of 1.3
and 1.5 respectively.

\vspace*{-0.4in}
Note that we have three scalars within 30 GeV of each other with decay widths of
$\sim 18$ GeV for the neutral and $\sim 21$ GeV for the charged scalar. They all
contribute to the excess and lead to the broader peak as seen in the CDF data.
We also get a good fit to the ratio between the events at the peak for scalars and
the SM gauge boson which comes out to be $\sim 0.46$.

An alternative way of generating the excess in $Wjj$ events would have been to use the couplings
$Y_u [\overline{c}_L u_R \Phi_{21}^0+ \overline{s}_L u_R \Phi_{21}^-]$ with the slightly less suppressed
strange quark flux \cite{nelson}. However, with $Y_u \simeq 1.4$, we find that the LO cross section
for $Wjj$ is in excess of 8 pb, which is too large when compared to the events observed by CDF.
We also note that one could explain the $Wjj$ excess by making the charged scalars heavier
so that there is a resonance enhancement in the $W^\pm\Phi^0_{12}$ production. However, CDF
data discards any parent resonance contributing to the excess, which would have been
otherwise visible in the $M_{\ell\nu jj}$ distribution.
The present CDF analysis on the $Wjj$ events have also ruled out any similar excess
in the $Zjj$ channel. The $Zjj$ contribution in our model is
suppressed by numerical factors and
due to destructive interference between the s-channel ($u\bar{c}\to \Phi^{0*}_{12}\to\Phi^0_{12} Z$) and
t-channel ($u\bar{c}\to\Phi^0_{12} Z$) contributions. Similar
cancelations occur for the charged scalar too leading to a total cross section of
$\simeq 0.17$ pb for $p\bar{p}\to Z jj$, which is small and hard to observe at CDF.

{\it Predictions of the model:} Having shown the consistency of the model with both anomalies
reported by CDF, we now turn to its predictions.  (i) We obtain significant cross section for
the production of the scalars in association with photon at Tevatron. Demanding some
basic selection cuts on the photon of $p_T>30$ GeV and $|\eta|<2.0$, we get $\sigma(\gamma
\Phi^0_{12})= 2.1$ pb and $\sigma(\gamma \Phi^\pm_{12})= 0.133$ pb. (ii) The
light scalars in our model will be produced with larger cross sections at LHC. We find
the LO cross section for the associated production of these scalars with SM electroweak gauge bosons
listed below:
{\small \begin{eqnarray}
\sigma(Z \Phi^0_{12})\simeq 2.8~ pb;~~~ \sigma( Z \Phi^\mp_{12} )
\simeq 3.3~pb;~~~ \sigma( \gamma \Phi^0_{12})\simeq 23.8~ pb;  \nonumber \\
\sigma(\gamma \Phi^\mp_{12} ) \simeq 3.3 ~ pb;~ \sigma(W^\pm \Phi^0_{12})\simeq 73~ pb;
~~~\sigma(W^\pm \Phi^\mp_{12} ) \simeq 86 ~ pb. \nonumber
\end{eqnarray}}
\noindent
The photon cross section is subjected to the same selection cuts
mentioned above for Tevatron. It is worth noting that the $W\Phi_{12}$
mode has a cross section in the range of $\sim 150$ pb which is
comparable to the $t\bar{t}$ cross section at LHC. Thus LHC should be
able to see these scalars in the lepton+dijet events soon, if they are
responsible for the $Wjj$ anomaly reported by CDF. (iii) As our model also
predicts additional scalars ($\Phi_{23}$) which couple to top quark and
are heavier than the top quark ($200-400$ GeV), we can produce these
scalars in association with $\Phi_{13}$ which can have some implication
on top quark physics at the LHC. We do not find any significant
contribution to $t\bar{t}$ production but the $\Phi_{13}^\pm$ and $\Phi_{23}$ produced
will decay to a top/anti-top quark and light jet with 100\% probability. We
find $\sigma(\Phi_{13}^0\Phi_{13}^\pm) \simeq 5.6$ pb which can
affect single top studies at LHC. For a lighter $m_{\Phi_{23}}=200$ GeV, we get
additional source for single top events with $\sigma(\Phi_{13}^0 \Phi_{23})\simeq 3.5$ pb
which when combined with $\Phi_{13}^\pm$ is about 10\% of the single top production in the SM at LHC.
(iv) The operator $|Y_u|^2 (\overline{d}_L c_R)(\overline{c}_R d_L)/M_{\Phi_{12}}^2$
induced by $\Phi_{12}^\pm$ exchange, when written in the mass eigenstate basis
of the quarks ($d_L \simeq d_L^0 +  |V_{us}| s_L^0+  |V_{ub}| b_L^0$, where
the quark mixing is assumed to arise entirely from the down sector)
will generate new contributions for e.g., to the decay $B \rightarrow J/\Psi K_S$, which can modify
the prediction for $\sin2\beta$ by about 10\% from its SM value.  Since our model resembles type II two Higgs
doublet model with $\tan\beta \simeq 1$, there is a lower limit on the charged Higgs from $\Phi_{33}^{u,d}$,
which is about 250 GeV from $R_b$ and $b \rightarrow s\gamma$ \cite{logan}.  No other large flavor
changing effects are expected.

We are grateful to A. Khanov and F. Rizatdinova for
helpful discussions. KSB and SKR are supported in part by the US Department of Energy, Grant
Numbers DE-FG02-04ER41306 and DE-FG02-04ER46140. MF acknowledges NSERC of Canada for partial financial support
and the OSU HEP group for hospitality.

\vspace{-0.2in}
\bibliography{CDF results}

\begin{thebibliography}{999}
\bibitem{fn} The relation $m_e/m_\mu/m_\tau = m_d/m_s/m_b$ will be corrected by one--loop
  diagrams.
\end{thebibliography}

\end{document}